\documentclass[sigconf]{acmart}

\AtBeginDocument{%
   }
 
\setcopyright{acmlicensed}
\copyrightyear{2025}
\acmYear{2025}
\setcopyright{cc}
\setcctype{by-nc-sa}
\acmConference[DIS '25 Companion]{Designing Interactive Systems
	Conference}{July 5--9, 2025}{Funchal, Portugal}
\acmBooktitle{Designing Interactive Systems Conference (DIS '25 Companion),July 5--9, 2025, Funchal, Portugal}\acmDOI{10.1145/3715668.3736365}
\acmISBN{979-8-4007-1486-3/2025/07}

\usepackage{enumitem}
\usepackage[utf8x]{inputenc}
\usepackage{graphicx}
 
\begin{document}
\title{Towards a Working Definition of Designing Generative User Interfaces}

\author{Kyungho Lee}
\affiliation{%
  \institution{Expressive Computing Lab. \\ Department of Design, UNIST}
  \city{Ulsan}
  \country{Republic of Korea}}
\email{kyungho@unist.ac.kr}
\renewcommand{\shortauthors}{Lee et al.}

\begin{abstract}
Generative UI is transforming interface design by facilitating AI-driven collaborative workflows between designers and computational systems. This study establishes a working definition of Generative UI through a multi-method qualitative approach, integrating insights from a systematic literature review of 127 publications, expert interviews with 18 participants, and analyses of 12 case studies. Our findings identify five core themes that position Generative UI as an iterative and co-creative process. We highlight emerging design models, including hybrid creation, curation-based workflows, and AI-assisted refinement strategies. Additionally, we examine ethical challenges, evaluation criteria, and interaction models that shape the field. By proposing a conceptual foundation, this study advances both theoretical discourse and practical implementation, guiding future HCI research toward responsible and effective generative UI design practices.

\end{abstract}

\ccsdesc[300]{Human-centered computing~Interface design prototyping}
\ccsdesc[300]{Human-centered computing~Natural language interfaces}
\ccsdesc[500]{Human-centered computing~Interactive systems and tools}

\keywords{Generative UI; User Interface Design; AI-driven Design; Generative Design; Computational Creativity; Responsible AI}
 
 
\maketitle
 
\section{Introduction}
The landscape of User Interface (UI) design is experiencing a profound transformation driven by recent advances in generative artificial intelligence (GenAI) \cite{goodfellow_etal_2014_Generativeadversarialnets}. Once the exclusive domain of human designers working with static design tools like Adobe Creative Suite \cite{AdobeCreativeCloud}, UI creation is increasingly becoming a collaborative process where AI systems actively generate, refine, and even evaluate design solutions. This shift is not merely an expansion of the designer’s toolset— it represents a fundamental reconfiguration of the design process itself, challenging long-established paradigms and notions of creativity, authorship, and human-computer collaboration in user interface (UI) design \cite{yang_etal_2020_ReexaminingWhetherWhyHowHumanAIInteractionUniquelyDifficultDesign,carroll_etal_1988_Interfacemetaphorsuserinterfacedesign,stone_etal_2005_Userinterfacedesignevaluation,blair-early_zender_2008_Userinterfacedesignprinciplesinteractiondesign}. In recent years, there has been an unprecedented increase in both research interest and commercial applications using generative capabilities for UI design. Large language models such as GPT-4 \cite{OpenAI_GPT4o} and Claude \cite{Claude_AI_New} have demonstrated the ability to generate functional GUI code and design specifications. These technologies have rapidly moved from research-level outputs to production tools, with AI assistants like GitHub Copilot \cite{GitHub_Copilot}, Cursor.ai \cite{Cursor_AI}, or Bolt.new \cite{Bolt_new} embedding generative capabilities directly into professional design workflows. The result is a growing ecosystem where designers are increasingly collaborating with GenAI systems that can generate, refine, and implement design solutions at unprecedented speed and scale, even called vibe coding \cite{ray2025review}.

Despite this rapid technological advancement and practical adoption, the conceptual understanding of \textit{"Generative UI (GenUI)"} and \textit{"the design practices of GenUIs"} remains underdeveloped. The current understandings of UI/UX issues for generative AI \cite{kim2024ui} or intelligent user interfaces \cite{alvarez2007current} focus on technological advancements to enable GenUI design practices, static artifact generation or inference mechanisms and algorithms to monitor user actions, interactions, offering limited insight into the design-time, co-creative, and dialogic processes that define and characterize GenUI. 
This theoretical gap creates potential limitations for researchers investigating generative UI design phenomena and practitioners seeking to develop effective workflows that incorporate these new capabilities under a unified concept and shared conceptual working definitions.

From this perspective, this position paper addresses a conceptual gap by proposing a working definition of GenUIs and the design of GenUIs as a distinct design paradigm. Based on a systematic synthesis of academic literature, practitioner insights, and case analyses, we identify five constitutive elements that differentiate GenUI from both traditional UI design: computational co-creation, expanded design space exploration, representation fluidity, contextual adaptation, and synthesis over selection. Our findings suggest that GenUI is a new mode of interface creation that emphasizes design-time collaboration between human and machine agents and run-time user interaction with AI-generated GUI variants.

By providing this working definition, we aim to support the DIS community's ongoing efforts to understand and effectively leverage the transformative potential of generative AI, particularly in interface design. We believe that this conceptual clarity will become increasingly significant as generative approaches transition from experimental applications to mainstream design practices, fundamentally altering how interfaces are created, experienced, and adapted in response to user needs.

\section{Related Work}
Regarding \textit{the use of Generative AI for Design}, GenAI represents a paradigm shift in computational systems, moving from analytical to creative capabilities. GenAI models learn from large datasets, identifying underlying patterns and structures. This allows the models to create new content that resembles the training data \cite{goodfellow_etal_2014_Generativeadversarialnets}. Also, attention mechanisms, which allow models to focus on relevant parts of the input data. This is crucial for generating coherent and contextually accurate outputs, especially in tasks involving sequences \cite{vaswani_etal_2017_Attentionallyouneed}.

With the characteristics, the application of GenAIs to design disciplines has evolved rapidly over the past decade. Early explorations focused on relatively constrained generative systems for specific design tasks, such as typography generation \cite{campbell_kautz_2014_Learningmanifoldfonts}, layout optimization \cite{odonovan_etal_2015_DesignScapeDesignInteractiveLayoutSuggestions}, and color palette suggestion \cite{kita_miyata_2016_AestheticRatingColorSuggestionColorPalettes}, rather than being applied to the design of user interface tasks. These systems typically operated within narrowly defined parameters and required substantial human supervision. More recent advances have dramatically expanded both the scope and autonomy of generative design systems, enabling them to produce complex, multi-faceted design artifacts with minimal human guidance
\cite{matejka_etal_2018_DreamLensExplorationVisualizationLargeScaleGenerativeDesignDatasets,chen_zhang_2019_Learningimplicitfieldsgenerativeshapemodeling,chen_etal_2020_Generativepretrainingpixels,ramesh_etal_2021_Zeroshottexttoimagegeneration}.

Interestingly, it is considered that the concept of \textit{Generative Design} precedes current GenAI implementations, with roots in various design traditions that emphasize procedural or rule-based creation. In architecture and industrial design, generative design has historically referred to approaches that use algorithms to explore design spaces constrained by specified parameters and objectives \cite{fischer_herr_2001_Teachinggenerativedesign}. For example, Oxman's taxonomy of digital design  \cite{oxman_2006_Theorydesignfirstdigitalage} distinguishes between parametric design (manipulation of variables within a fixed system), algorithmic design (rule-based procedures), and truly generative approaches where emergent properties arise from system interactions.

In creative fields, in a broad manner, generative methodologies have been conceptualized through various theoretical frameworks. For instance, Gero's Function-Behavior-Structure model \cite{gero_kannengiesser_2004_situatedfunctionbehaviourstructureframework} provides a lens for understanding how generative systems navigate from functional requirements to structural design solutions. Cross' designerly ways of knowing \cite{cross_1982_Designerlywaysknowing} emphasizes the distinct epistemological approaches that characterize design thinking, many of which present challenges for computational implementation. Dorst's frame creation \cite{dorst_2011_coredesignthinkinganditsapplication} offers insights into how designers reframe problems—a critical creative capability that current generative systems struggle to replicate.

Computational creativity research offers additional perspectives relevant to generative UI. Boden's conceptualization of creativity as exploratory, combinatorial, or transformational \cite{boden_2004_creativemindMythsmechanisms} provides a useful framework for evaluating the creative capabilities of generative UI systems. Colton's creative tripod model \cite{colton_2008_CreativityPerceptionCreativityComputationalSystems}, emphasizing skill, appreciation, and imagination as essential components of computational creativity, suggests criteria for assessing the sophistication of generative design tools.
Within the HCI research, several theoretical frameworks inform understanding of generative approaches. Reflection-in-action, as articulated by Schön \cite{schon_2017_reflectivepractitionerHowprofessionalsthinkaction}, captures the iterative, conversational nature of design processes that effective generative UI systems must support. Participatory design traditions \cite{muller_kuhn_1993_Participatorydesign} raise important questions about how generative tools redistribute agency between designers, systems, and end-users. User-centered design methodologies \cite{gasson_2003_Humancenteredvsusercenteredapproachesinformationsystemdesign,anderson_1988_Usercenteredsystemdesignnewperspectiveshumancomputerinteraction} emphasize empirical user research as a foundation for design decisions—an aspect often underdeveloped in current generative approaches.

Building on these insights, our research draws from foundational concepts to explore how GenUI systems function as generative, dialogic partners in the design process. By integrating these theoretical foundations with empirical analyses of expert interviews and case studies, we develop a working definition and dimensional framework for GenUI that clarifies how it differs from prior user interface design paradigms. The proposed framework aims to guide future inquiries into how generative tools reshape interface design at both conceptual and operational levels.

\section{Identifying Gaps in Literature}
\begin{figure*}[!h!t]
	\centering
	\includegraphics[width=\linewidth]{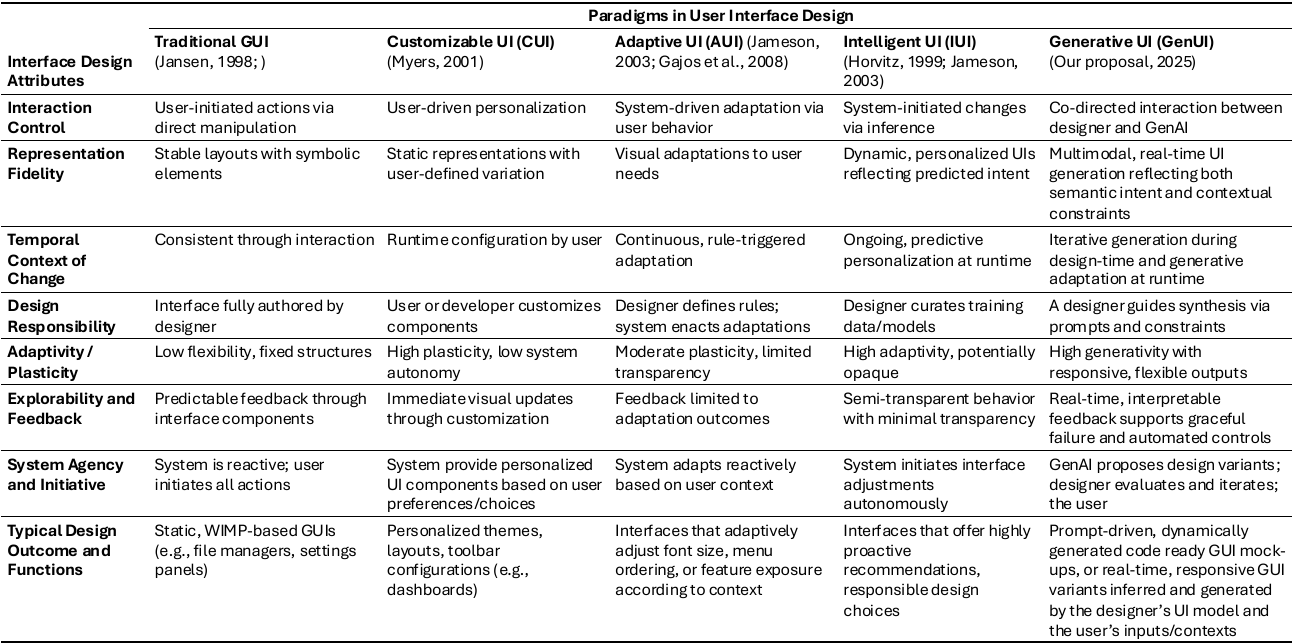}
	\caption{A summary of user interface design paradigms based on design attributes \cite{jansen1998graphical,carroll1988interface,myers2001using,jameson2007adaptive,horvitz1999principles} }
	\label{fig:paradigms}
\end{figure*}
Despite this rich theoretical landscape, existing definitions and frameworks potentially fall short in addressing the unique characteristics of generative UI design. Current conceptualizations of generative design could not specifically fit into to account for several critical aspects of UI-specific applications as follows:
First, the interactive nature of UI artifacts introduces unique challenges and opportunities not present in static design domains. Unlike generative art or architectural form-finding, UI generation must consider not only visual aesthetics but also interaction flows, state transitions, and real-time responsiveness to user input \cite{blom_beckhaus_2014_designspacedynamicinteractivevirtualenvironments,altakrouri_schrader_2012_dynamicnaturalinteractionensembles}. 

Current generative design frameworks rarely address these dynamic aspects adequately.
Second, UI design operates within particularly complex constraint ecosystems, including technical implementation limitations, accessibility requirements, cross-platform compatibility, and evolving design systems \cite{zhu_etal_2018_ValueSensitiveAlgorithmDesignMethodCaseStudyLessons}. Generative approaches must navigate these constraints while maintaining creative flexibility.
Third, the collaborative nature of contemporary UI design processes, involving multiple stakeholders with diverse expertise, creates distinct requirements for generative tools \cite{inie_dalsgaard_2020_HowInteractionDesignersUseToolsManageIdeas}. Unlike domains where a single designer can work with generative systems, for instance, graphic design or illustration, UI design typically involves cross-functional teams whose collaborative dynamics are significantly altered by the introduction of generative capabilities.
Finally, the rapid evolution of GenUI technologies has created a disconnection between theoretical understanding and practical implementation. Industry adoption has outpaced academic conceptualization, resulting in a terminology gap where practitioners and researchers lack a shared vocabulary to discuss emerging phenomena \cite{yang_etal_2019_UnremarkableAIFittingIntelligentDecisionSupportCriticalClinicalDecisionMakingProcesses}.

To clarify the conceptual distinctiveness of GenUI, we present a comparative framework situating GenUI within the historical evolution of interface paradigms—from traditional graphical user interfaces to customizable, adaptive, and intelligent systems as shown in \autoref{fig:paradigms}. 
This chronological structure provides a historical narrative: from early symbolic GUIs (e.g., WIMP interfaces) to user-configurable systems, then to system-adaptive and intelligent behaviors, and finally to co-creative design-time systems. It offers a theoretical ground that GenUI represents a paradigm shift in how interfaces are designed, not just how they operate, enabled by technological advances through conceptual dimensions. 

The interface design attributes used in \autoref{fig:paradigms} to differentiate user interface paradigms are grounded in well-established HCI theory. For example, interaction control draws from Norman’s action cycle \cite{norman1986cognitive}, which models how users form goals, execute actions, and evaluate outcomes, thus framing how control is distributed between human and system. Representation fidelity builds on Shneiderman’s \cite{shneiderman1983direct,shneiderman2003eyes} visual information-seeking mantra and Nielsen’s \cite{nielsen1994usability} and Whiteside et al.'s \cite{whiteside1988usability} heuristic of system visibility, emphasizing the importance of interfaces that transparently reflect internal states. Dix et al.’s interaction framework informs temporal context \cite{dix2010human}, distinguishing between design-time and runtime processes in user interaction. Design responsibility is situated within Fischer and Giaccardi’s \cite{fischer2004meta} meta-design paradigm, which examines how co-creation distributes creative authority across human and machine agents. Adaptivity and plasticity emerge from Thevenin and Coutaz’s \cite{thevenin1999plasticity} work on user interface plasticity and Myers’s \cite{myers1998brief} historical analysis of flexible UI technologies. Finally, explorability and feedback are rooted in Nielsen’s \cite{nielsen1994usability} emphasis on user control and Shneiderman’s \cite{shneiderman1983direct} advocacy for direct manipulation. At the same time, system agency and initiative are extended from Jameson’s \cite{jameson2007adaptive} and Horvitz’s \cite{horvitz1999principles} models of proactive, intelligent interfaces. 

Together, these dimensions offer a theoretical lens for understanding the unique position of GenUI within the evolution of interface paradigms - GenUI is not just oriented toward generating design artifacts but also alters user experience during runtime according to user preferences and contexts. The dimensions, such as interaction control, temporal logic, and system agency, help us to better understand how these systems operate at runtime and what they produce at design time.

\section{Research Methods}
We employed a multi-method qualitative approach to develop a working definition of generative UI, integrating theoretical perspectives with emerging practices. Our research consisted of four components: a systematic literature review, expert interviews, case study analysis, and community discourse.

First, we conducted a systematic literature review using an adapted PRISMA protocol \cite{moher2010preferred}. Our objective was to trace how generative approaches have been conceptualized and applied within the context of UI design, and to distinguish GenUI from historical paradigms such as \texttt{adaptive}, \texttt{customizable}, \texttt{personalizable}, \texttt{intelligent},\texttt{ai-assisted}, and \texttt{generative} interfaces. We searched major digital databases—including the ACM Digital Library, IEEE Xplore, and Google Scholar—using search terms: combining \texttt{historical paradigms keyword} + \texttt{relevant keywords} as suffix (user interface, design, interaction, system) respectively. Our inclusion criteria focused on publications from 1982 (in 1982 CHI was organised for the first time) to December 2024 to capture both foundational user interface concepts and recent advances in generative AI for interface design.

Our initial search yielded 1,213 records. After removing duplicates and screening titles and abstracts for relevance, we excluded 932 records that did not directly address interface or user interaction design. The exclusion criteria applied were as follows: (1) papers primarily focused on developing algorithms rather than design; (2) papers that described generative systems in less relevant domains (e.g., architecture, art) without addressing interactive systems; and (3) papers primarily focused on building generative AI models without a design-oriented framework/intepretation. A remaining set of 281 papers underwent a full-text review.

Lastly, we conducted an open coding process on the final corpus of 127 papers. Initial codes were clustered inductively into conceptual categories, including design agency, representational mechanisms, system initiative, iteration logic, and contextual constraint handling. Through iterative refinement through triangulation, we synthesized these categories into five core dimensions differentiating GenUI: computational co-creation, expanded design space exploration, representation fluidity, contextual adaptation, and synthesis over selection. These dimensions informed the construction of our working definition and comparative framework, anchoring our contribution within both empirical insights and theoretical lineage in HCI and design studies.

To complement the systematic literature review results with practitioner perspectives, we conducted semi-structured interviews with 18 experts: HCI researchers at a university (n=6), UI/UX designers using generative tools (n=7), and generative UI developers (n=5). Participants, aged 27–48 (M=36.4, SD=7.8), had an average of 3.3 years of experience with generative AI. Interviews (18.6 total hours in total, 8,167 words in each interview) explored definitions, implementation experiences, benefits, limitations, and future expectations, transcribed and analyzed using our coding framework. 

Additionally, we analyzed 12 case studies from participants’ former/current projects, covering experimental research systems (8), AI-assisted mobile app prototypes, and enterprise platforms (2) for image content generation (2). We examined how generative capabilities were conceptualized, implemented, and integrated into professional UI design workflows. Data analysis, conducted in ATLAS.ti, used affinity diagramming to identify themes and conceptual relationships. To enhance validity, we conducted member checking with eight interviewees, incorporating their feedback into the final definition. The methodological approach described above enabled us to develop a working definition that reflects both theoretical rigor and practical relevance. By synthesizing perspectives from literature, expert practitioners, implemented systems, and community discourse, we aimed to create a definition with broad applicability across research and practice contexts within the evolving field of generative UI design.

\section{Findings}
\subsection{The Working Definition of Designing Generative UI}
Our systematic analysis revealed several key themes that collectively characterize the emerging domain of generative user interface design. We propose the following working definition of a \textit{Generative User Interface}: it is a user interface that is created, refined, or evolved through computational generative processes, where AI systems contribute to the design, implementation, or adaptation of the interface. This approach ensures that the interface accurately captures, analyzes, and implements user values, resulting in interfaces that are not only technically functional but also meaningfully aligned with users' actual values as suggested by  \cite{zhu_etal_2018_ValueSensitiveAlgorithmDesignMethodCaseStudyLessons}. These interfaces emerge from collaborative creation between human designers, AI assist tools and agents, and the final users, rather than being solely manually crafted through traditional design process and deliver to the final users.

Under this context, \textit{Designing Generative UI} refers to a computational co-creation process in which designers and AI systems collaboratively explore, synthesize, and refine user interfaces. This is achieved through bidirectional translation between representational forms, context-aware pattern generation, and iterative evaluation against human and technical requirements. This approach enhances the design process by facilitating collaboration between humans and computational agents (AI systems), thereby expanding the design spaces that can be explored in the latent space, which is informed by data. Additionally, it allows for rapid adaptation to evolving situated design problems in a non-linear manner. This definition encompasses both the technological components of generative UI—such as computational pattern generation and multi-modal representation—and the socio-technical processes through which these capabilities are implemented, including co-creation, distributed agency, and iterative evaluation.

\begin{itemize}
	\item \textbf{Computational Co-Creation} :Generative UI fosters a human-AI partnership rather than full automation. Unlike earlier visions of design automation that positioned computational systems as eventually replacing human designers, current understanding emphasizes human-AI partnership in what several interview participants described as "human-AI co-creation" or "augmented design intelligence." As one expert UI designer explained, "\textit{It's not about the AI taking over design work, but rather establishing a creative dialogue. I provide direction, constraints, and critical evaluation, while the generative system offers multiple variations, possibilities, and sometimes surprising alternatives I wouldn't have considered.}" (P9, Senior Designer)
	\item \textbf{Expanded Design Space Exploration} : another consistent theme was generative UI's capacity to facilitate exploration of expanded design spaces in complex possibility spaces as suggested \cite{boden_2004_creativemindMythsmechanisms}. Traditional UI design processes often converge prematurely on familiar solutions due to time constraints, cognitive biases, or limited design vocabulary as noted [31]. Generative approaches enable broader exploration by rapidly producing diverse alternatives, including options that might fall outside a designer's typical patterns. This was particularly emphasized in case study examples, where systems like Uizard's UI generation features or Cursor.ai assistant tools were  positioned as tools for expanding creative possibilities rather than merely accelerating production of predetermined designs. One research publication characterized this as "computational divergent thinking" \cite{krish_2011_practicalgenerativedesignmethod}, suggesting that this computational variations couldn't replace human creativity as we still evaluate its quality and relevance to the project but amplifies it by suggesting unexpected directions. 	
	\item \textbf{Representation Fluidity} : Unlike linear UI design workflows, generative UI allows seamless movement between sketches, wireframes, visual designs, and functional code. Tools like Uizard and Galileo AI facilitate this multi-modal design synthesis, accelerating such iteration. This bidirectional translation between representations emerged as a distinctively generative capability, contrasting with traditional tools that typically operate within a single representational domain. As one system developer noted: \textit{"What makes current generative UI truly revolutionary is this ability to jump between ideation, written product requirement document texts, and code almost instantaneously. A designer can describe an interaction pattern in words, immediately see visual alternatives, refine through direct manipulation, and then export functional code—all within minutes instead of days."} (P15, AI Tool Developer)
	\item \textbf{Contextual Adaptation} : Generative UI adapts to brand guidelines, accessibility standards, user preferences, and cultural factors, demonstrating an implicit design knowledge that refines outputs dynamically beyond explicit rules. One research paper characterized this as "implicit design knowledge" that enables generative systems to make nuanced contextual adjustments that would be difficult to explicitly codify \cite{becker_etal_2021_Reconstructingimplicitknowledgelanguagemodels}. One of the interviewees described: "\textit{I've built a collection of prompt templates with the blank for different design scenarios—one for creating initial explorations of a new feature, another for generating variations on an existing component, others for adapting designs across different design issues. Each template includes specific language and expressions that I've found effectively communicates my design intent to the AI.}" (P7, Product Designer)
	\item \textbf{Design Synthesis Over Selection} : The final theme involved the shift from selection-based to synthesis-based design processes. Traditional digital design tools primarily offer selection from predefined design options or with specific user scenario for designers to create genuinely novel elements from the scratch. Generative UI approaches, by contrast, enable continuous synthesis of new design possibilities through recombination and transformation embracing various design issues. As one GUI designer explained: "\textit{Before, my design process felt like IKEA shopping — selecting from available options and assembling them. With generative tools, it's more like cooking — I'm combining ingredients with a general recipe in mind with my own kicks, but I know the specific outcome emerges through the process. I'm still making creative decisions, but they're higher-level directional choices rather than selections from limited menus.}" (P3, UX Director)
	\item \textbf{Gaps in Runtime Adaptation to User Values} :  While GenUI systems possess the technical ability to synthesize interface variants based on high-level parameters—such as device constraints, accessibility requirements, or use-case descriptors—these capabilities are often underutilized or externally specified, rather than embedded into the generative logic itself. Prompt engineering and iterative refinement remain centered on designer intent, aesthetic variation, and structural efficiency, but rarely incorporate dynamic user needs or runtime context (e.g., task urgency, preferred user values, contextual constraints) as first-class design considerations so far. This creates a disconnect between the AI system’s generative power and the situated nature of user interaction, ultimately limiting the alignment between generative outputs and end-user goals and values. For example, "\textit{I know the GenAI system can generate a bunch of variants in my end, but I never thought about how to get it to reflect what the user needs and what s/he wants to see in the real-time context. It still feels like I am utilizing a traditional design and deployment process.}" (P3, UX Director)
\end{itemize}

\subsection{Challenges Ahead}
Our research identified key challenges and open questions in the design and the use of generative UIs, spanning ethical, technological, and professional concerns. These issues demand further exploration by the HCI community to shape the responsible evolution of this field.
\begin{itemize}
	\item \textbf{Value prioritization and balance} : The generative approach creates a more direct pathway from value articulation to design implementation, reducing the "value drift" that often occurs in traditional design processes where initial insights become diluted through multiple translation stages. Also, by formally extracting and representing values, GenUI systems should be able to effectively handle complex value trade-offs, recognizing when certain values represent boundary constraints while others are optimization targets within those boundaries.
	\item  \textbf{Bias and Representation} : Generative systems learn from historical design data, inheriting and potentially amplifying biases related to gender, culture, and accessibility. Without intervention, these biases may persist in AI-generated outputs. 
	\item  \textbf{Attribution and Intellectual Property} : Generative UI challenges traditional notions of authorship and ownership. When AI plays a central role in creation, who receives credit—the designer, the AI, or both? Many practitioners struggle to showcase their contributions when AI handles much of the visible production
	\item  \textbf{Functional Accuracy in Interactive Contexts} : Unlike static UI designs, GenUIs should be able to function correctly across states and user flows. Current generative models often create interfaces that look correct but fail in usability or interaction logic. 
	\item  \textbf{Integration with Existing Design Workflows} : Most generative tools remain siloed, requiring context-switching between AI-driven interfaces and traditional design platforms being serviced. Seamless integration and transition is essential for broader adoption but remains technically challenging.
\end{itemize}

\subsection{Future Research Direction}
\subsubsection{Designing Together: Supporting Collaboration Between Humans and Generative Systems}
The findings synthesized through our research reveal that GenUI reconfigures traditional boundaries of interaction control, design responsibility, and system agency. Unlike earlier paradigms—where systems were either passive instruments (GUI, customizable UI) or reactive entities (AUI/IUI)—GenUI tools engage in bidirectional exchanges with designers. Our analysis shows that these tools do not merely execute instructions but interpret, propose, and adapt design outputs in response to evolving human input. However, while this suggests a shift toward shared authorship, existing frameworks lack a clear model for managing such co-agency in real-world workflows. Thus, future research should build collaborative frameworks that formalize how prompts, feedback loops, and system-generated alternatives function as joint acts of design reasoning. These frameworks should map the interdependencies between human intent and machine proposal, clarifying how each party contributes to and negotiates the evolving artifact.

\subsubsection{Changing How We Design: Understanding the Impact of GenUI on User Interface Design Practice}
Our literture review identifies representation fidelity, temporal context, and adaptivity/plasticity as dimensions through which GenUI transforms traditional design practices. GenUI enables seamless transitions between text, image, code, and layout at design time, facilitating real-time synthesis rather than post hoc adaptation. This affordance dramatically expands the exploratory potential of interface design but also risks displacing critical evaluation with rapid iteration. Future work must investigate how these capabilities affect both the quality and originality of design outcomes over time. For instance, does multimodal translation encourage broader ideation or reinforce superficial convergence on familiar patterns? How do designers use—or misuse—adaptive generation during high-stakes decision-making? By examining how GenUI tools affect representation choices, iteration rhythms, and solution diversity, future research can assess the creative and procedural impacts of generative assistance across settings.

\subsubsection{Building Trustworthy Tools: Addressing Ethical and Responsible Use of GenUIs}
The feedback/explorability, design responsibility, and system agency dimensions in our findings point to critical ethical and epistemological challenges. While GenUI systems afford rapid feedback and ideation through dialogic interaction, they often mask the underlying generative logic or rationale behind their creations. Designers may struggle to understand why certain suggestions were made, whether outputs are contextually appropriate, or how the input influences generative outomces. This creates a gap between surface usability and deep accountability. Moreover, the shifting locus of authorship—where systems contribute substantively to the interface’s structure—raises unresolved questions about intellectual ownership, accountability for biased outputs, and user agency. Future work must embed mechanisms for transparency, contestability, and reconfigurability into GenUI workflows. This includes features such as provenance tracking, ethical auditing of training data, and value-aligned prompt structures to ensure that generative tools serve diverse and inclusive design values, not just efficiency or novelty.

\section{Conclusion}
This paper introduced a working definition of designing Generative User Interfaces (GenUI), clarifying key characteristics that set GenUI apart from traditional UI development and other forms of generative design. Through a systematic literature review, expert interviews, and case studies, we identified five foundational themes. These themes frame GenUI as an interactive, socio-technical design process, where human designers and AI systems iteratively co-construct interface artifacts. The definition offers a shared vocabulary and conceptual grounding for analyzing and designing generative systems in UI contexts. Beyond conceptual clarification, this work provides practical insights for understanding GenUI in terms of design outcomes as well as design process, emphasizing hybrid creation workflows, curation-based roles, and iterative refinement strategies.

While this study offers a foundational perspective, several limitations must be noted. The empirical base—though rich in qualitative insights—draws primarily from early adoption contexts and professional design environments, potentially limiting generalizability to non-professional or cross-domain settings. Our expert sample, while diverse, may skew toward frontier tool adopters, underrepresenting conventional practices. Additionally, the interpretive nature of our thematic synthesis involves subjective judgment, particularly in delineating evolving constructs like “co-creation.” Future work should complement these findings with longitudinal studies, experimental evaluations, and participatory methods to validate and extend the framework proposed here.

Looking ahead, several research directions remain critical. First, new evaluation frameworks are needed to assess GenUI outputs beyond usability, incorporating novelty, contextual fit, transparency, and ethical alignment. Second, optimal models of human–AI interaction require exploration: How should systems visualize intermediate states, and how can feedback mechanisms best support interpretability? Third, GenUI is reshaping design cognition, necessitating studies on how generative tools influence problem-solving and learning. Fourth, maintaining user-centeredness remains a challenge, as generative systems risk prioritizing visual output over real user needs. Finally, generative capabilities may enable novel methodologies such as adaptive prototyping, dynamic personalization, and continuous interface evolution—rich areas for future inquiry. As generative technologies continue to evolve, so must the framework, which adapts to emerging modalities, capabilities, and user values in a more responsible manner.

\begin{acks}
I am grateful to my collaborators, such as Dr. Dajung Kim, Seoyeong Hwang and Haeeun Shin, for their thoughtful feedback, critical insights, and generous engagement throughout this work. Their perspectives enriched the depth and direction of the research, particularly in clarifying the framework and polishing the findings. This research was partially supported by a grant from the Korea Institute for Advancement of Technology (KIAT) funded by the Government of Korea (MOTIE) (P0025495, Establishment of Infrastructure for Integrated Utilization of Design Industry Data). This work was also partially supported by the Technology Innovation Program (20015056, Commercialization design and development of Intelligent Product-Service System for personalized full silver life cycle care) funded by the Ministry of Trade, Industry \& Energy(MOTIE, Korea).
\end{acks}

\bibliographystyle{ACM-Reference-Format}
\bibliography{dis25companion-86}


\begin{thebibliography}{55}


\ifx \showCODEN    \undefined \def \showCODEN     #1{\unskip}     \fi
\ifx \showISBNx    \undefined \def \showISBNx     #1{\unskip}     \fi
\ifx \showISBNxiii \undefined \def \showISBNxiii  #1{\unskip}     \fi
\ifx \showISSN     \undefined \def \showISSN      #1{\unskip}     \fi
\ifx \showLCCN     \undefined \def \showLCCN      #1{\unskip}     \fi
\ifx \shownote     \undefined \def \shownote      #1{#1}          \fi
\ifx \showarticletitle \undefined \def \showarticletitle #1{#1}   \fi
\ifx \showURL      \undefined \def \showURL       {\relax}        \fi
\providecommand\bibfield[2]{#2}
\providecommand\bibinfo[2]{#2}
\providecommand\natexlab[1]{#1}
\providecommand\showeprint[2][]{arXiv:#2}

\bibitem[AI(2025)]%
        {Claude_AI_New}
\bibfield{author}{\bibinfo{person}{Claude AI}.}
  \bibinfo{year}{2025}\natexlab{}.
\newblock \bibinfo{title}{New Claude AI}.
\newblock
\urldef\tempurl%
\url{https://claude.ai/new}
\showURL{%
\tempurl}
\newblock
\shownote{Accessed: 2025-03-22}.


\bibitem[Altakrouri and Schrader(2012)]%
        {altakrouri_schrader_2012_dynamicnaturalinteractionensembles}
\bibfield{author}{\bibinfo{person}{Bashar Altakrouri} {and}
  \bibinfo{person}{Andreas Schrader}.} \bibinfo{year}{2012}\natexlab{}.
\newblock \showarticletitle{Towards dynamic natural interaction ensembles}. In
  \bibinfo{booktitle}{\emph{the 26th {BCS} {Conference} on {Human} {Computer}
  {Interaction}}}. \bibinfo{publisher}{BCS Learning \& Development}.
\newblock
\urldef\tempurl%
\url{https://www.scienceopen.com/hosted-document?doi=10.14236/ewic/HCI2012.85}
\showURL{%
\tempurl}


\bibitem[Alvarez-Cortes et~al\mbox{.}(2007)]%
        {alvarez2007current}
\bibfield{author}{\bibinfo{person}{Victor Alvarez-Cortes},
  \bibinfo{person}{Benjamin~E Zayas-Perez}, \bibinfo{person}{Victor~Huga
  Zarate-Silva}, {and} \bibinfo{person}{Jorge A~Ramirez Uresti}.}
  \bibinfo{year}{2007}\natexlab{}.
\newblock \showarticletitle{Current trends in adaptive user interfaces:
  Challenges and applications}. In \bibinfo{booktitle}{\emph{Electronics,
  Robotics and Automotive Mechanics Conference (CERMA 2007)}}. IEEE,
  \bibinfo{pages}{312--317}.
\newblock


\bibitem[Anderson(1988)]%
  {anderson_1988_Usercenteredsystemdesignnewperspectiveshumancomputerinteraction}
\bibfield{author}{\bibinfo{person}{Nancy~S. Anderson}.}
  \bibinfo{year}{1988}\natexlab{}.
\newblock \bibinfo{title}{User centered system design: new perspectives on
  human-computer interaction}.
\newblock
\urldef\tempurl%
\url{https://www.jstor.org/stable/1422802}
\showURL{%
\tempurl}


\bibitem[Becker et~al\mbox{.}(2021)]%
        {becker_etal_2021_Reconstructingimplicitknowledgelanguagemodels}
\bibfield{author}{\bibinfo{person}{Maria Becker}, \bibinfo{person}{Siting
  Liang}, {and} \bibinfo{person}{Anette Frank}.}
  \bibinfo{year}{2021}\natexlab{}.
\newblock \showarticletitle{Reconstructing implicit knowledge with language
  models}. In \bibinfo{booktitle}{\emph{Proceedings of {Deep} {Learning}
  {Inside} {Out} ({DeeLIO}): {The} 2nd {Workshop} on {Knowledge} {Extraction}
  and {Integration} for {Deep} {Learning} {Architectures}}}.
  \bibinfo{pages}{11--24}.
\newblock
\urldef\tempurl%
\url{https://aclanthology.org/2021.deelio-1.2/}
\showURL{%
\tempurl}


\bibitem[Blair-Early and Zender(2008)]%
  {blair-early_zender_2008_Userinterfacedesignprinciplesinteractiondesign}
\bibfield{author}{\bibinfo{person}{Adream Blair-Early} {and}
  \bibinfo{person}{Mike Zender}.} \bibinfo{year}{2008}\natexlab{}.
\newblock \showarticletitle{User interface design principles for interaction
  design}.
\newblock \bibinfo{journal}{\emph{Design Issues}} \bibinfo{volume}{24},
  \bibinfo{number}{3} (\bibinfo{year}{2008}), \bibinfo{pages}{85--107}.
\newblock
\urldef\tempurl%
\url{https://www.jstor.org/stable/25224185}
\showURL{%
\tempurl}
\newblock
\shownote{Publisher: JSTOR}.


\bibitem[Blom and Beckhaus(2014)]%
        {blom_beckhaus_2014_designspacedynamicinteractivevirtualenvironments}
\bibfield{author}{\bibinfo{person}{Kristopher~J. Blom} {and}
  \bibinfo{person}{Steffi Beckhaus}.} \bibinfo{year}{2014}\natexlab{}.
\newblock \showarticletitle{The design space of dynamic interactive virtual
  environments}.
\newblock \bibinfo{journal}{\emph{Virtual Reality}} \bibinfo{volume}{18},
  \bibinfo{number}{2} (\bibinfo{date}{June} \bibinfo{year}{2014}),
  \bibinfo{pages}{101--116}.
\newblock
\showISSN{1359-4338, 1434-9957}
\href{https://doi.org/10.1007/s10055-013-0232-y}{doi:\nolinkurl{10.1007/s10055-013-0232-y}}


\bibitem[Boden(2004)]%
        {boden_2004_creativemindMythsmechanisms}
\bibfield{author}{\bibinfo{person}{Margaret~A. Boden}.}
  \bibinfo{year}{2004}\natexlab{}.
\newblock \bibinfo{booktitle}{\emph{The creative mind: {Myths} and
  mechanisms}}.
\newblock \bibinfo{publisher}{Routledge}.
\newblock


\bibitem[Bolt(2025)]%
        {Bolt_new}
\bibfield{author}{\bibinfo{person}{Bolt}.} \bibinfo{year}{2025}\natexlab{}.
\newblock \bibinfo{title}{Bolt New}.
\newblock
\urldef\tempurl%
\url{https://bolt.new/}
\showURL{%
\tempurl}
\newblock
\shownote{Accessed: 2025-03-22}.


\bibitem[Campbell and Kautz(2014)]%
        {campbell_kautz_2014_Learningmanifoldfonts}
\bibfield{author}{\bibinfo{person}{Neill D.~F. Campbell} {and}
  \bibinfo{person}{Jan Kautz}.} \bibinfo{year}{2014}\natexlab{}.
\newblock \showarticletitle{Learning a manifold of fonts}.
\newblock \bibinfo{journal}{\emph{ACM Transactions on Graphics}}
  \bibinfo{volume}{33}, \bibinfo{number}{4} (\bibinfo{date}{July}
  \bibinfo{year}{2014}), \bibinfo{pages}{1--11}.
\newblock
\showISSN{0730-0301, 1557-7368}
\href{https://doi.org/10.1145/2601097.2601212}{doi:\nolinkurl{10.1145/2601097.2601212}}


\bibitem[Carroll et~al\mbox{.}(1988a)]%
        {carroll_etal_1988_Interfacemetaphorsuserinterfacedesign}
\bibfield{author}{\bibinfo{person}{John~M. Carroll}, \bibinfo{person}{Robert~L.
  Mack}, {and} \bibinfo{person}{Wendy~A. Kellogg}.}
  \bibinfo{year}{1988}\natexlab{a}.
\newblock \showarticletitle{Interface metaphors and user interface design}.
\newblock In \bibinfo{booktitle}{\emph{Handbook of human-computer
  interaction}}. \bibinfo{publisher}{Elsevier}, \bibinfo{pages}{67--85}.
\newblock
\urldef\tempurl%
\url{https://www.sciencedirect.com/science/article/pii/B9780444705365500087}
\showURL{%
\tempurl}


\bibitem[Carroll et~al\mbox{.}(1988b)]%
        {carroll1988interface}
\bibfield{author}{\bibinfo{person}{John~M Carroll}, \bibinfo{person}{Robert~L
  Mack}, {and} \bibinfo{person}{Wendy~A Kellogg}.}
  \bibinfo{year}{1988}\natexlab{b}.
\newblock \showarticletitle{Interface metaphors and user interface design}.
\newblock In \bibinfo{booktitle}{\emph{Handbook of human-computer
  interaction}}. \bibinfo{publisher}{Elsevier}, \bibinfo{pages}{67--85}.
\newblock


\bibitem[Chen et~al\mbox{.}(2020)]%
        {chen_etal_2020_Generativepretrainingpixels}
\bibfield{author}{\bibinfo{person}{Mark Chen}, \bibinfo{person}{Alec Radford},
  \bibinfo{person}{Rewon Child}, \bibinfo{person}{Jeffrey Wu},
  \bibinfo{person}{Heewoo Jun}, \bibinfo{person}{David Luan}, {and}
  \bibinfo{person}{Ilya Sutskever}.} \bibinfo{year}{2020}\natexlab{}.
\newblock \showarticletitle{Generative pretraining from pixels}. In
  \bibinfo{booktitle}{\emph{International conference on machine learning}}.
  \bibinfo{publisher}{PMLR}, \bibinfo{pages}{1691--1703}.
\newblock
\urldef\tempurl%
\url{http://proceedings.mlr.press/v119/chen20s.html}
\showURL{%
\tempurl}


\bibitem[Chen and Zhang(2019)]%
        {chen_zhang_2019_Learningimplicitfieldsgenerativeshapemodeling}
\bibfield{author}{\bibinfo{person}{Zhiqin Chen} {and} \bibinfo{person}{Hao
  Zhang}.} \bibinfo{year}{2019}\natexlab{}.
\newblock \showarticletitle{Learning implicit fields for generative shape
  modeling}. In \bibinfo{booktitle}{\emph{Proceedings of the {IEEE}/{CVF}
  conference on computer vision and pattern recognition}}.
  \bibinfo{pages}{5939--5948}.
\newblock


\bibitem[Colton(2008)]%
        {colton_2008_CreativityPerceptionCreativityComputationalSystems}
\bibfield{author}{\bibinfo{person}{Simon Colton}.}
  \bibinfo{year}{2008}\natexlab{}.
\newblock \showarticletitle{Creativity {Versus} the {Perception} of
  {Creativity} in {Computational} {Systems}.}. In
  \bibinfo{booktitle}{\emph{{AAAI} spring symposium: creative intelligent
  systems}}, Vol.~\bibinfo{volume}{8}. \bibinfo{publisher}{Palo Alto, CA},
  \bibinfo{pages}{7}.
\newblock


\bibitem[Cross(1982)]%
        {cross_1982_Designerlywaysknowing}
\bibfield{author}{\bibinfo{person}{Nigel Cross}.}
  \bibinfo{year}{1982}\natexlab{}.
\newblock \showarticletitle{Designerly ways of knowing}.
\newblock \bibinfo{journal}{\emph{Design studies}} \bibinfo{volume}{3},
  \bibinfo{number}{4} (\bibinfo{year}{1982}), \bibinfo{pages}{221--227}.
\newblock
\urldef\tempurl%
\url{https://www.sciencedirect.com/science/article/pii/0142694X82900400}
\showURL{%
\tempurl}
\newblock
\shownote{Publisher: Elsevier}.


\bibitem[Cursor(2025)]%
        {Cursor_AI}
\bibfield{author}{\bibinfo{person}{Cursor}.} \bibinfo{year}{2025}\natexlab{}.
\newblock \bibinfo{title}{Cursor AI}.
\newblock
\urldef\tempurl%
\url{https://www.cursor.com/}
\showURL{%
\tempurl}
\newblock
\shownote{Accessed: 2025-03-22}.


\bibitem[Dix(2010)]%
        {dix2010human}
\bibfield{author}{\bibinfo{person}{Alan Dix}.} \bibinfo{year}{2010}\natexlab{}.
\newblock \showarticletitle{Human--computer interaction: A stable discipline, a
  nascent science, and the growth of the long tail}.
\newblock \bibinfo{journal}{\emph{Interacting with computers}}
  \bibinfo{volume}{22}, \bibinfo{number}{1} (\bibinfo{year}{2010}),
  \bibinfo{pages}{13--27}.
\newblock


\bibitem[Dorst(2011)]%
        {dorst_2011_coredesignthinkinganditsapplication}
\bibfield{author}{\bibinfo{person}{Kees Dorst}.}
  \bibinfo{year}{2011}\natexlab{}.
\newblock \showarticletitle{The core of `design thinking'and its application}.
\newblock \bibinfo{journal}{\emph{Design studies}} \bibinfo{volume}{32},
  \bibinfo{number}{6} (\bibinfo{year}{2011}), \bibinfo{pages}{521--532}.
\newblock
\urldef\tempurl%
\url{https://www.sciencedirect.com/science/article/pii/S0142694X11000603}
\showURL{%
\tempurl}
\newblock
\shownote{Publisher: Elsevier}.


\bibitem[Fischer et~al\mbox{.}(2004)]%
        {fischer2004meta}
\bibfield{author}{\bibinfo{person}{Gerhard Fischer}, \bibinfo{person}{Elisa
  Giaccardi}, \bibinfo{person}{Yunwen Ye}, \bibinfo{person}{Alistair~G
  Sutcliffe}, {and} \bibinfo{person}{Nikolay Mehandjiev}.}
  \bibinfo{year}{2004}\natexlab{}.
\newblock \showarticletitle{Meta-design: a manifesto for end-user development}.
\newblock \bibinfo{journal}{\emph{Commun. ACM}} \bibinfo{volume}{47},
  \bibinfo{number}{9} (\bibinfo{year}{2004}), \bibinfo{pages}{33--37}.
\newblock


\bibitem[Fischer and Herr(2001)]%
        {fischer_herr_2001_Teachinggenerativedesign}
\bibfield{author}{\bibinfo{person}{Thomas Fischer} {and}
  \bibinfo{person}{Christiane~M. Herr}.} \bibinfo{year}{2001}\natexlab{}.
\newblock \showarticletitle{Teaching generative design}. In
  \bibinfo{booktitle}{\emph{Proceedings of the 4th {Conference} on {Generative}
  {Art}}}. \bibinfo{publisher}{Politechnico di Milano University Milan},
  \bibinfo{pages}{8}.
\newblock
\urldef\tempurl%
\url{http://www.generativeart.com/on/cic/ga2001_pdf/fischer.pdf}
\showURL{%
\tempurl}


\bibitem[Gasson(2003)]%
  {gasson_2003_Humancenteredvsusercenteredapproachesinformationsystemdesign}
\bibfield{author}{\bibinfo{person}{Susan Gasson}.}
  \bibinfo{year}{2003}\natexlab{}.
\newblock \showarticletitle{Human-centered vs. user-centered approaches to
  information system design}.
\newblock \bibinfo{journal}{\emph{Journal of Information Technology Theory and
  Application (JITTA)}} \bibinfo{volume}{5}, \bibinfo{number}{2}
  (\bibinfo{year}{2003}), \bibinfo{pages}{5}.
\newblock
\urldef\tempurl%
\url{https://aisel.aisnet.org/cgi/viewcontent.cgi?article=1153&context=jitta}
\showURL{%
\tempurl}


\bibitem[Gero and Kannengiesser(2004)]%
        {gero_kannengiesser_2004_situatedfunctionbehaviourstructureframework}
\bibfield{author}{\bibinfo{person}{John~S. Gero} {and} \bibinfo{person}{Udo
  Kannengiesser}.} \bibinfo{year}{2004}\natexlab{}.
\newblock \showarticletitle{The situated function--behaviour--structure
  framework}.
\newblock \bibinfo{journal}{\emph{Design studies}} \bibinfo{volume}{25},
  \bibinfo{number}{4} (\bibinfo{year}{2004}), \bibinfo{pages}{373--391}.
\newblock
\urldef\tempurl%
\url{https://www.sciencedirect.com/science/article/pii/S0142694X03000735}
\showURL{%
\tempurl}
\newblock
\shownote{Publisher: Elsevier}.


\bibitem[GitHub(2025)]%
        {GitHub_Copilot}
\bibfield{author}{\bibinfo{person}{GitHub}.} \bibinfo{year}{2025}\natexlab{}.
\newblock \bibinfo{title}{GitHub Copilot}.
\newblock
\urldef\tempurl%
\url{https://github.com/features/copilot}
\showURL{%
\tempurl}
\newblock
\shownote{Accessed: 2025-03-22}.


\bibitem[Goodfellow et~al\mbox{.}(2014)]%
        {goodfellow_etal_2014_Generativeadversarialnets}
\bibfield{author}{\bibinfo{person}{Ian~J. Goodfellow}, \bibinfo{person}{Jean
  Pouget-Abadie}, \bibinfo{person}{Mehdi Mirza}, \bibinfo{person}{Bing Xu},
  \bibinfo{person}{David Warde-Farley}, \bibinfo{person}{Sherjil Ozair},
  \bibinfo{person}{Aaron Courville}, {and} \bibinfo{person}{Yoshua Bengio}.}
  \bibinfo{year}{2014}\natexlab{}.
\newblock \showarticletitle{Generative adversarial nets}.
\newblock \bibinfo{journal}{\emph{Advances in neural information processing
  systems}}  \bibinfo{volume}{27} (\bibinfo{year}{2014}).
\newblock
\urldef\tempurl%
\url{https://proceedings.neurips.cc/paper_files/paper/2014/hash/f033ed80deb0234979a61f95710dbe25-Abstract.html}
\showURL{%
\tempurl}


\bibitem[Horvitz(1999)]%
        {horvitz1999principles}
\bibfield{author}{\bibinfo{person}{Eric Horvitz}.}
  \bibinfo{year}{1999}\natexlab{}.
\newblock \showarticletitle{Principles of mixed-initiative user interfaces}. In
  \bibinfo{booktitle}{\emph{Proceedings of the SIGCHI conference on Human
  Factors in Computing Systems}}. \bibinfo{pages}{159--166}.
\newblock


\bibitem[Inc.(2025)]%
        {AdobeCreativeCloud}
\bibfield{author}{\bibinfo{person}{Adobe Inc.}}
  \bibinfo{year}{2025}\natexlab{}.
\newblock \bibinfo{title}{Adobe Creative Cloud}.
\newblock
\urldef\tempurl%
\url{https://www.adobe.com/creativecloud.html}
\showURL{%
\tempurl}
\newblock
\shownote{Accessed: 2025-03-22}.


\bibitem[Inie and Dalsgaard(2020)]%
        {inie_dalsgaard_2020_HowInteractionDesignersUseToolsManageIdeas}
\bibfield{author}{\bibinfo{person}{Nanna Inie} {and} \bibinfo{person}{Peter
  Dalsgaard}.} \bibinfo{year}{2020}\natexlab{}.
\newblock \showarticletitle{How {Interaction} {Designers} {Use} {Tools} to
  {Manage} {Ideas}}.
\newblock \bibinfo{journal}{\emph{ACM Transactions on Computer-Human
  Interaction}} \bibinfo{volume}{27}, \bibinfo{number}{2}
  (\bibinfo{date}{April} \bibinfo{year}{2020}), \bibinfo{pages}{1--26}.
\newblock
\showISSN{1073-0516, 1557-7325}
\href{https://doi.org/10.1145/3365104}{doi:\nolinkurl{10.1145/3365104}}


\bibitem[Jameson(2007)]%
        {jameson2007adaptive}
\bibfield{author}{\bibinfo{person}{Anthony Jameson}.}
  \bibinfo{year}{2007}\natexlab{}.
\newblock \showarticletitle{Adaptive interfaces and agents}.
\newblock In \bibinfo{booktitle}{\emph{The human-computer interaction
  handbook}}. \bibinfo{publisher}{CRC press}, \bibinfo{pages}{459--484}.
\newblock


\bibitem[Jansen(1998)]%
        {jansen1998graphical}
\bibfield{author}{\bibinfo{person}{Bernard~J Jansen}.}
  \bibinfo{year}{1998}\natexlab{}.
\newblock \showarticletitle{The graphical user interface}.
\newblock \bibinfo{journal}{\emph{ACM SIGCHI Bulletin}} \bibinfo{volume}{30},
  \bibinfo{number}{2} (\bibinfo{year}{1998}), \bibinfo{pages}{22--26}.
\newblock


\bibitem[Kim et~al\mbox{.}(2024)]%
        {kim2024ui}
\bibfield{author}{\bibinfo{person}{Tae-seok Kim}, \bibinfo{person}{Marvin~John
  Ignacio}, \bibinfo{person}{Seounghee Yu}, \bibinfo{person}{Hulin Jin}, {and}
  \bibinfo{person}{Yong-guk Kim}.} \bibinfo{year}{2024}\natexlab{}.
\newblock \showarticletitle{UI/UX for Generative AI: Taxonomy, Trend, and
  Challenge}.
\newblock \bibinfo{journal}{\emph{IEEE Access}} (\bibinfo{year}{2024}).
\newblock


\bibitem[Kita and Miyata(2016)]%
        {kita_miyata_2016_AestheticRatingColorSuggestionColorPalettes}
\bibfield{author}{\bibinfo{person}{N. Kita} {and} \bibinfo{person}{K. Miyata}.}
  \bibinfo{year}{2016}\natexlab{}.
\newblock \showarticletitle{Aesthetic {Rating} and {Color} {Suggestion} for
  {Color} {Palettes}}.
\newblock \bibinfo{journal}{\emph{Computer Graphics Forum}}
  \bibinfo{volume}{35}, \bibinfo{number}{7} (\bibinfo{date}{Oct.}
  \bibinfo{year}{2016}), \bibinfo{pages}{127--136}.
\newblock
\showISSN{0167-7055, 1467-8659}
\href{https://doi.org/10.1111/cgf.13010}{doi:\nolinkurl{10.1111/cgf.13010}}


\bibitem[Krish(2011)]%
        {krish_2011_practicalgenerativedesignmethod}
\bibfield{author}{\bibinfo{person}{Sivam Krish}.}
  \bibinfo{year}{2011}\natexlab{}.
\newblock \showarticletitle{A practical generative design method}.
\newblock \bibinfo{journal}{\emph{Computer-aided design}} \bibinfo{volume}{43},
  \bibinfo{number}{1} (\bibinfo{year}{2011}), \bibinfo{pages}{88--100}.
\newblock
\urldef\tempurl%
\url{https://www.sciencedirect.com/science/article/pii/S0010448510001764}
\showURL{%
\tempurl}
\newblock
\shownote{Publisher: Elsevier}.


\bibitem[Matejka et~al\mbox{.}(2018)]%
  {matejka_etal_2018_DreamLensExplorationVisualizationLargeScaleGenerativeDesignDatasets}
\bibfield{author}{\bibinfo{person}{Justin Matejka}, \bibinfo{person}{Michael
  Glueck}, \bibinfo{person}{Erin Bradner}, \bibinfo{person}{Ali Hashemi},
  \bibinfo{person}{Tovi Grossman}, {and} \bibinfo{person}{George Fitzmaurice}.}
  \bibinfo{year}{2018}\natexlab{}.
\newblock \showarticletitle{Dream {Lens}: {Exploration} and {Visualization} of
  {Large}-{Scale} {Generative} {Design} {Datasets}}. In
  \bibinfo{booktitle}{\emph{Proceedings of the 2018 {CHI} {Conference} on
  {Human} {Factors} in {Computing} {Systems}}}. \bibinfo{publisher}{ACM},
  \bibinfo{address}{Montreal QC Canada}, \bibinfo{pages}{1--12}.
\newblock
\showISBNx{978-1-4503-5620-6}
\href{https://doi.org/10.1145/3173574.3173943}{doi:\nolinkurl{10.1145/3173574.3173943}}


\bibitem[Moher et~al\mbox{.}(2010)]%
        {moher2010preferred}
\bibfield{author}{\bibinfo{person}{David Moher}, \bibinfo{person}{Alessandro
  Liberati}, \bibinfo{person}{Jennifer Tetzlaff}, \bibinfo{person}{Douglas~G
  Altman}, \bibinfo{person}{Prisma Group}, {et~al\mbox{.}}}
  \bibinfo{year}{2010}\natexlab{}.
\newblock \showarticletitle{Preferred reporting items for systematic reviews
  and meta-analyses: the PRISMA statement}.
\newblock \bibinfo{journal}{\emph{International journal of surgery}}
  \bibinfo{volume}{8}, \bibinfo{number}{5} (\bibinfo{year}{2010}),
  \bibinfo{pages}{336--341}.
\newblock


\bibitem[Muller and Kuhn(1993)]%
        {muller_kuhn_1993_Participatorydesign}
\bibfield{author}{\bibinfo{person}{Michael~J. Muller} {and}
  \bibinfo{person}{Sarah Kuhn}.} \bibinfo{year}{1993}\natexlab{}.
\newblock \showarticletitle{Participatory design}.
\newblock \bibinfo{journal}{\emph{Commun. ACM}} \bibinfo{volume}{36},
  \bibinfo{number}{6} (\bibinfo{date}{June} \bibinfo{year}{1993}),
  \bibinfo{pages}{24--28}.
\newblock
\showISSN{0001-0782, 1557-7317}
\href{https://doi.org/10.1145/153571.255960}{doi:\nolinkurl{10.1145/153571.255960}}


\bibitem[Myers(1998)]%
        {myers1998brief}
\bibfield{author}{\bibinfo{person}{Brad~A Myers}.}
  \bibinfo{year}{1998}\natexlab{}.
\newblock \showarticletitle{A brief history of human-computer interaction
  technology}.
\newblock \bibinfo{journal}{\emph{interactions}} \bibinfo{volume}{5},
  \bibinfo{number}{2} (\bibinfo{year}{1998}), \bibinfo{pages}{44--54}.
\newblock


\bibitem[Myers(2001)]%
        {myers2001using}
\bibfield{author}{\bibinfo{person}{Brad~A Myers}.}
  \bibinfo{year}{2001}\natexlab{}.
\newblock \showarticletitle{Using handhelds and PCs together}.
\newblock \bibinfo{journal}{\emph{Commun. ACM}} \bibinfo{volume}{44},
  \bibinfo{number}{11} (\bibinfo{year}{2001}), \bibinfo{pages}{34--41}.
\newblock


\bibitem[Nielsen(1994)]%
        {nielsen1994usability}
\bibfield{author}{\bibinfo{person}{Jakob Nielsen}.}
  \bibinfo{year}{1994}\natexlab{}.
\newblock \bibinfo{booktitle}{\emph{Usability engineering}}.
\newblock \bibinfo{publisher}{Morgan Kaufmann}.
\newblock


\bibitem[Norman(1986)]%
        {norman1986cognitive}
\bibfield{author}{\bibinfo{person}{Donald~A Norman}.}
  \bibinfo{year}{1986}\natexlab{}.
\newblock \showarticletitle{Cognitive engineering}.
\newblock \bibinfo{journal}{\emph{User centered system design}}
  \bibinfo{volume}{31}, \bibinfo{number}{61} (\bibinfo{year}{1986}),
  \bibinfo{pages}{2}.
\newblock


\bibitem[O'Donovan et~al\mbox{.}(2015)]%
        {odonovan_etal_2015_DesignScapeDesignInteractiveLayoutSuggestions}
\bibfield{author}{\bibinfo{person}{Peter O'Donovan}, \bibinfo{person}{Aseem
  Agarwala}, {and} \bibinfo{person}{Aaron Hertzmann}.}
  \bibinfo{year}{2015}\natexlab{}.
\newblock \showarticletitle{{DesignScape}: {Design} with {Interactive} {Layout}
  {Suggestions}}. In \bibinfo{booktitle}{\emph{Proceedings of the 33rd {Annual}
  {ACM} {Conference} on {Human} {Factors} in {Computing} {Systems}}}.
  \bibinfo{publisher}{ACM}, \bibinfo{address}{Seoul Republic of Korea},
  \bibinfo{pages}{1221--1224}.
\newblock
\showISBNx{978-1-4503-3145-6}
\href{https://doi.org/10.1145/2702123.2702149}{doi:\nolinkurl{10.1145/2702123.2702149}}


\bibitem[OpenAI(2025)]%
        {OpenAI_GPT4o}
\bibfield{author}{\bibinfo{person}{OpenAI}.} \bibinfo{year}{2025}\natexlab{}.
\newblock \bibinfo{title}{Hello GPT-4o}.
\newblock
\urldef\tempurl%
\url{https://openai.com/index/hello-gpt-4o/}
\showURL{%
\tempurl}
\newblock
\shownote{Accessed: 2025-03-22}.


\bibitem[Oxman(2006)]%
        {oxman_2006_Theorydesignfirstdigitalage}
\bibfield{author}{\bibinfo{person}{Rivka Oxman}.}
  \bibinfo{year}{2006}\natexlab{}.
\newblock \showarticletitle{Theory and design in the first digital age}.
\newblock \bibinfo{journal}{\emph{Design studies}} \bibinfo{volume}{27},
  \bibinfo{number}{3} (\bibinfo{year}{2006}), \bibinfo{pages}{229--265}.
\newblock
\urldef\tempurl%
\url{https://www.sciencedirect.com/science/article/pii/S0142694X05000840}
\showURL{%
\tempurl}
\newblock
\shownote{Publisher: Elsevier}.


\bibitem[Ramesh et~al\mbox{.}(2021)]%
        {ramesh_etal_2021_Zeroshottexttoimagegeneration}
\bibfield{author}{\bibinfo{person}{Aditya Ramesh}, \bibinfo{person}{Mikhail
  Pavlov}, \bibinfo{person}{Gabriel Goh}, \bibinfo{person}{Scott Gray},
  \bibinfo{person}{Chelsea Voss}, \bibinfo{person}{Alec Radford},
  \bibinfo{person}{Mark Chen}, {and} \bibinfo{person}{Ilya Sutskever}.}
  \bibinfo{year}{2021}\natexlab{}.
\newblock \showarticletitle{Zero-shot text-to-image generation}. In
  \bibinfo{booktitle}{\emph{International conference on machine learning}}.
  \bibinfo{publisher}{Pmlr}, \bibinfo{pages}{8821--8831}.
\newblock
\urldef\tempurl%
\url{https://proceedings.mlr.press/v139/ramesh21a.html?ref=journey}
\showURL{%
\tempurl}


\bibitem[Ray(2025)]%
        {ray2025review}
\bibfield{author}{\bibinfo{person}{Partha~Pratim Ray}.}
  \bibinfo{year}{2025}\natexlab{}.
\newblock \showarticletitle{A Review on Vibe Coding: Fundamentals,
  State-of-the-art, Challenges and Future Directions}.
\newblock  (\bibinfo{year}{2025}).
\newblock


\bibitem[Sch{\"o}n(2017)]%
        {schon_2017_reflectivepractitionerHowprofessionalsthinkaction}
\bibfield{author}{\bibinfo{person}{Donald~A. Sch{\"o}n}.}
  \bibinfo{year}{2017}\natexlab{}.
\newblock \bibinfo{booktitle}{\emph{The reflective practitioner: {How}
  professionals think in action}}.
\newblock \bibinfo{publisher}{Routledge}.
\newblock
\urldef\tempurl%
\url{https://www.taylorfrancis.com/books/mono/10.4324/9781315237473/reflective-practitioner-donald-sch%C3%B6n}
\showURL{%
\tempurl}


\bibitem[Shneiderman(1983)]%
        {shneiderman1983direct}
\bibfield{author}{\bibinfo{person}{Ben Shneiderman}.}
  \bibinfo{year}{1983}\natexlab{}.
\newblock \showarticletitle{Direct manipulation: A step beyond programming
  languages}.
\newblock \bibinfo{journal}{\emph{Computer}} \bibinfo{volume}{16},
  \bibinfo{number}{08} (\bibinfo{year}{1983}), \bibinfo{pages}{57--69}.
\newblock


\bibitem[Shneiderman(2003)]%
        {shneiderman2003eyes}
\bibfield{author}{\bibinfo{person}{Ben Shneiderman}.}
  \bibinfo{year}{2003}\natexlab{}.
\newblock \showarticletitle{The eyes have it: A task by data type taxonomy for
  information visualizations}.
\newblock In \bibinfo{booktitle}{\emph{The craft of information
  visualization}}. \bibinfo{publisher}{Elsevier}, \bibinfo{pages}{364--371}.
\newblock


\bibitem[Stone et~al\mbox{.}(2005)]%
        {stone_etal_2005_Userinterfacedesignevaluation}
\bibfield{author}{\bibinfo{person}{Debbie Stone}, \bibinfo{person}{Caroline
  Jarrett}, \bibinfo{person}{Mark Woodroffe}, {and} \bibinfo{person}{Shailey
  Minocha}.} \bibinfo{year}{2005}\natexlab{}.
\newblock \bibinfo{booktitle}{\emph{User interface design and evaluation}}.
\newblock \bibinfo{publisher}{Elsevier}.
\newblock
\urldef\tempurl%
\url{https://books.google.com/books?hl=en&lr=&id=VvSoyqPBPbMC&oi=fnd&pg=PR21&dq=user+interface+design&ots=d9MWTXlLO8&sig=5g0t7oP5jDTGRi8zvspvF8IiotQ}
\showURL{%
\tempurl}


\bibitem[Thevenin and Coutaz(1999)]%
        {thevenin1999plasticity}
\bibfield{author}{\bibinfo{person}{David Thevenin} {and}
  \bibinfo{person}{Jo{\"e}lle Coutaz}.} \bibinfo{year}{1999}\natexlab{}.
\newblock \showarticletitle{Plasticity of user interfaces: Framework and
  research agenda.}. In \bibinfo{booktitle}{\emph{Interact}},
  Vol.~\bibinfo{volume}{99}. Citeseer, \bibinfo{pages}{110--117}.
\newblock


\bibitem[Vaswani et~al\mbox{.}(2017)]%
        {vaswani_etal_2017_Attentionallyouneed}
\bibfield{author}{\bibinfo{person}{Ashish Vaswani}, \bibinfo{person}{Noam
  Shazeer}, \bibinfo{person}{Niki Parmar}, \bibinfo{person}{Jakob Uszkoreit},
  \bibinfo{person}{Llion Jones}, \bibinfo{person}{Aidan~N. Gomez},
  \bibinfo{person}{{\L}ukasz Kaiser}, {and} \bibinfo{person}{Illia
  Polosukhin}.} \bibinfo{year}{2017}\natexlab{}.
\newblock \showarticletitle{Attention is all you need}.
\newblock \bibinfo{journal}{\emph{Advances in neural information processing
  systems}}  \bibinfo{volume}{30} (\bibinfo{year}{2017}).
\newblock
\urldef\tempurl%
\url{https://proceedings.neurips.cc/paper/2017/hash/3f5ee243547dee91fbd053c1c4a845aa-Abstract.html}
\showURL{%
\tempurl}


\bibitem[Whiteside et~al\mbox{.}(1988)]%
        {whiteside1988usability}
\bibfield{author}{\bibinfo{person}{John Whiteside}, \bibinfo{person}{John
  Bennett}, {and} \bibinfo{person}{Karen Holtzblatt}.}
  \bibinfo{year}{1988}\natexlab{}.
\newblock \showarticletitle{Usability engineering: Our experience and
  evolution}.
\newblock In \bibinfo{booktitle}{\emph{Handbook of human-computer
  interaction}}. \bibinfo{publisher}{Elsevier}, \bibinfo{pages}{791--817}.
\newblock


\bibitem[Yang et~al\mbox{.}(2020)]%
  {yang_etal_2020_ReexaminingWhetherWhyHowHumanAIInteractionUniquelyDifficultDesign}
\bibfield{author}{\bibinfo{person}{Qian Yang}, \bibinfo{person}{Aaron
  Steinfeld}, \bibinfo{person}{Carolyn Ros{\'e}}, {and} \bibinfo{person}{John
  Zimmerman}.} \bibinfo{year}{2020}\natexlab{}.
\newblock \showarticletitle{Re-examining {Whether}, {Why}, and {How}
  {Human}-{AI} {Interaction} {Is} {Uniquely} {Difficult} to {Design}}. In
  \bibinfo{booktitle}{\emph{Proceedings of the 2020 {CHI} {Conference} on
  {Human} {Factors} in {Computing} {Systems}}}. \bibinfo{publisher}{ACM},
  \bibinfo{address}{Honolulu HI USA}, \bibinfo{pages}{1--13}.
\newblock
\showISBNx{978-1-4503-6708-0}
\href{https://doi.org/10.1145/3313831.3376301}{doi:\nolinkurl{10.1145/3313831.3376301}}


\bibitem[Yang et~al\mbox{.}(2019)]%
  {yang_etal_2019_UnremarkableAIFittingIntelligentDecisionSupportCriticalClinicalDecisionMakingProcesses}
\bibfield{author}{\bibinfo{person}{Qian Yang}, \bibinfo{person}{Aaron
  Steinfeld}, {and} \bibinfo{person}{John Zimmerman}.}
  \bibinfo{year}{2019}\natexlab{}.
\newblock \showarticletitle{Unremarkable {AI}: {Fitting} {Intelligent}
  {Decision} {Support} into {Critical}, {Clinical} {Decision}-{Making}
  {Processes}}. In \bibinfo{booktitle}{\emph{Proceedings of the 2019 {CHI}
  {Conference} on {Human} {Factors} in {Computing} {Systems}}}.
  \bibinfo{publisher}{ACM}, \bibinfo{address}{Glasgow Scotland Uk},
  \bibinfo{pages}{1--11}.
\newblock
\showISBNx{978-1-4503-5970-2}
\href{https://doi.org/10.1145/3290605.3300468}{doi:\nolinkurl{10.1145/3290605.3300468}}


\bibitem[Zhu et~al\mbox{.}(2018)]%
        {zhu_etal_2018_ValueSensitiveAlgorithmDesignMethodCaseStudyLessons}
\bibfield{author}{\bibinfo{person}{Haiyi Zhu}, \bibinfo{person}{Bowen Yu},
  \bibinfo{person}{Aaron Halfaker}, {and} \bibinfo{person}{Loren Terveen}.}
  \bibinfo{year}{2018}\natexlab{}.
\newblock \showarticletitle{Value-{Sensitive} {Algorithm} {Design}: {Method},
  {Case} {Study}, and {Lessons}}.
\newblock \bibinfo{journal}{\emph{Proceedings of the ACM on Human-Computer
  Interaction}} \bibinfo{volume}{2}, \bibinfo{number}{CSCW}
  (\bibinfo{date}{Nov.} \bibinfo{year}{2018}), \bibinfo{pages}{1--23}.
\newblock
\showISSN{2573-0142}
\href{https://doi.org/10.1145/3274463}{doi:\nolinkurl{10.1145/3274463}}


\end{thebibliography}

\end{document}